\documentstyle[psfig,raulbib,times]{mn}
\title{Primordial Proto--Galaxies}
\author[Paolo Padoan, Raul Jimenez \& James S. Dunlop]{Paolo Padoan$^1$,
Raul Jimenez$^2$ \& James S. Dunlop$^2$ \\
$^1$Harvard University Department of Astronomy, Cambridge, MA 02138
(ppadoan@cfa.harvard.edu) \\
$^2$Institute for Astronomy, University of Edinburgh, Blackford Hill,
Edinburgh EH9 3HJ,Scotland, UK (raul@roe.ac.uk, jsd@roe.ac.uk)}

\begin{document}
\maketitle

\begin{abstract}
The first stars to form in protogalaxies must have primordial chemical
composition. We refer to a protogalaxy that is forming stars of primordial
composition, or very low metallicity ($Z\le 0.01 Z_{\odot}$), as ``Primordial
Proto--Galaxy'' (PPG). PPGs contain little or no dust, and therefore their
spectral energy distribution can be modelled from the rest--frame ultraviolet
to the infrared without accounting for dust extinction and emission. We
present the results of computing the photometric properties of high redshift
PPGs at near--to--mid infrared wavelengths, that will soon be available with
the new generation of infrared space telescopes, such as SIRTF and the
NGST. We show that: i) PPGs at very high redshift ($5<z<10$) should be easily
selected from deep near/mid IR surveys with a colour--colour criterion; ii)
PPGs at redshift $5<z<10$ can be detected at 8~$\mu$m with the NGST, if they
have a constant star formation rate of at least $100$~M$_{\odot}/$yr; iii)
once the redshift of a PPG photometric candidate is determined, its
near--to--mid infrared colors should provide strong constraints on the stellar
IMF at zero or very low metallicity.
\end{abstract}


\section{Introduction}

Photometric redshifts have been used very successfully in the last few years
to select high redshift galaxies ($z>2$) detected with deep surveys such as
the HDF (e.g. \scite{D+98} and ref. therein). Star forming galaxies have been
discovered out to redshift $z=6.7$ (\pcite{CLP99}). At $z>5$, traditional
optical bands (U, B, V, R) fall below the rest frame wavelength that
corresponds to the Lyman break spectral feature (1200 \AA), where most of the
stellar radiation is extinguished either by interstellar or intergalactic
hydrogen. Because of this, galaxies at $z>5$ are practically invisible at
those photometric bands, and even if they were detected, their colors would
provide very little information about their stellar population. New detectors
and space telescopes, such as SIRTF, and in particular the NGST, will offer the
possibility of detecting very distant galaxies at IR wavelengths, and of using
photometric redshifts also with far--IR colors, as proposed by \scite{SE99}.

Nearby star forming galaxies are known to contain a considerable amount of
dust, that extinguishes a large fraction of their stellar UV light and boosts,
even by orders of magnitude, their IR luminosity. The effect of dust must be
taken into account in the computation of colors involving photometric bands
that span a large wavelength interval. However, the search for very young
proto--galaxies, perhaps the first star forming systems in the Universe, might
be pursued without having to model the effects of dust. The first stars to
form in proto--galaxies must have primordial chemical composition, or at least
very low metallicity (e.g. \pcite{PJJ97}). Although it is difficult to model
the formation and disruption of dust grains on a galactic scale, it is likely
that the dust content of a galaxy grows together with its metallicity, and
that a proto--galaxy with primordial chemical composition or very low
metallicity ($Z\le 0.01 Z_{\odot}$) has practically no dust
(\pcite{JPDBJM99}).  In this work we refer to young protogalaxies with
metallicity $Z\le 0.01 Z_{\odot}$ as primordial protogalaxies, or PPGs.

In the next section we compute the IR color evolution with redshift of PPGs
with $Z=0.01 Z_{\odot}$, and show that PPGs at redshift $5<z<10$ have IR
colors similar to those of nearby ($z<0.5$) young stellar populations with
metallicity slightly under the solar value and with no dust. However, nearby
galaxies are either forming stars and contain dust (spiral, irregular and
starburst galaxies), or have lost most gas and dust and are not forming stars
anymore (elliptical galaxies), and in both cases they are even redder than
high redshift PPGs.  In \S 3 we show that, because of their very blue colour,
PPGs do not suffer from very strong cosmological dimming, a feature which
should aid their detectability in deep infrared surveys with the NGST. If
nearby galaxies ($z<0.5$), with infrared colors similar to PPGs, did exist
(though very rare), they could be distinguished from PPGs because they would
be spatially resolved.

Star forming gas of primordial composition is likely to be warmer than
present-day molecular gas, because it is cooled mainly by molecular hydrogen,
to temperatures probably in excess of 100~K. This relatively warm temperature 
is likely to affect the stellar initial mass function (IMF). In \S 4 we show
that, once the redshift of a PPG photometric candidate is determined,
near-to-mid IR colors can provide constraints on the stellar IMF. For example,
a stellar IMF containing as many stars with mass $M<5M_{\odot}$ as the solar
neighborhood IMF could potentially be readily excluded by the mid-IR colors,
in favor of an IMF with no stars with mass $M<5M_{\odot}$.
In \S 5 we briefly discuss the performance of the broad band selection
of PPGs, versus emission line searches, and show that the broad band
photometry is probably convenient. Our conclusions are summarized in \S 6.

\section{Spectral energy distribution and IR colors of PPGs}

One of the most difficult problems in the study of star forming galaxies at
high redshift is to quantify the effect of dust upon their spectral energy
distribution (SED).  The largely unknown effect of dust results into a very
uncertain conversion from UV and (sub--)mm luminosities to star formation
rates (e.g. \scite{P+98,JPDBJM99,Peacock+99}).  Uncertainties in the
temperature, geometry, and amount of dust in high redshift galaxies are the
main sources of errors in the estimate of star formation rates in the high
redshift Universe (\pcite{JPDBJM99}).

Traditional near infrared band colors (e.g. $J-K$) of galaxies at redshift
$2<z<4$ are not strongly affected by dust extinction (at most 0.5 mag in
reasonable models -- \scite{JPDBJM99}). In contrast, colors of high-redshift
objects derived from combining photometric bands spanning a larger range of
wavelengths can be much more severely affected by the presence of dust, both
due to dust extinction in the rest-frame UV, and dust emission in the
rest--frame IR and near--IR.  However, very young proto--galaxies of
primordial chemical compositions, or at least very low metallicity ($Z\le 0.01
Z_{\odot}$), are likely to have little or no dust. Because of the absence of
dust and of the extremely low metallicity (and perhaps of their ``massive''
stellar IMF --see \S 4), PPGs are the rest--frame bluest stellar systems in
the Universe. In fact, in the redshift interval $5<z<10$, they have colours
similar to the colours of nearby ($z<0.5$) idealized stabursts with no dust
and slightly sub--solar metallicity. They are definitely bluer than nearby
real starbursts and elliptical galaxies, and even bluer than nearby star
forming disc and irregular galaxies (see Figure~2).

\begin{figure}
\centerline{ \psfig{figure=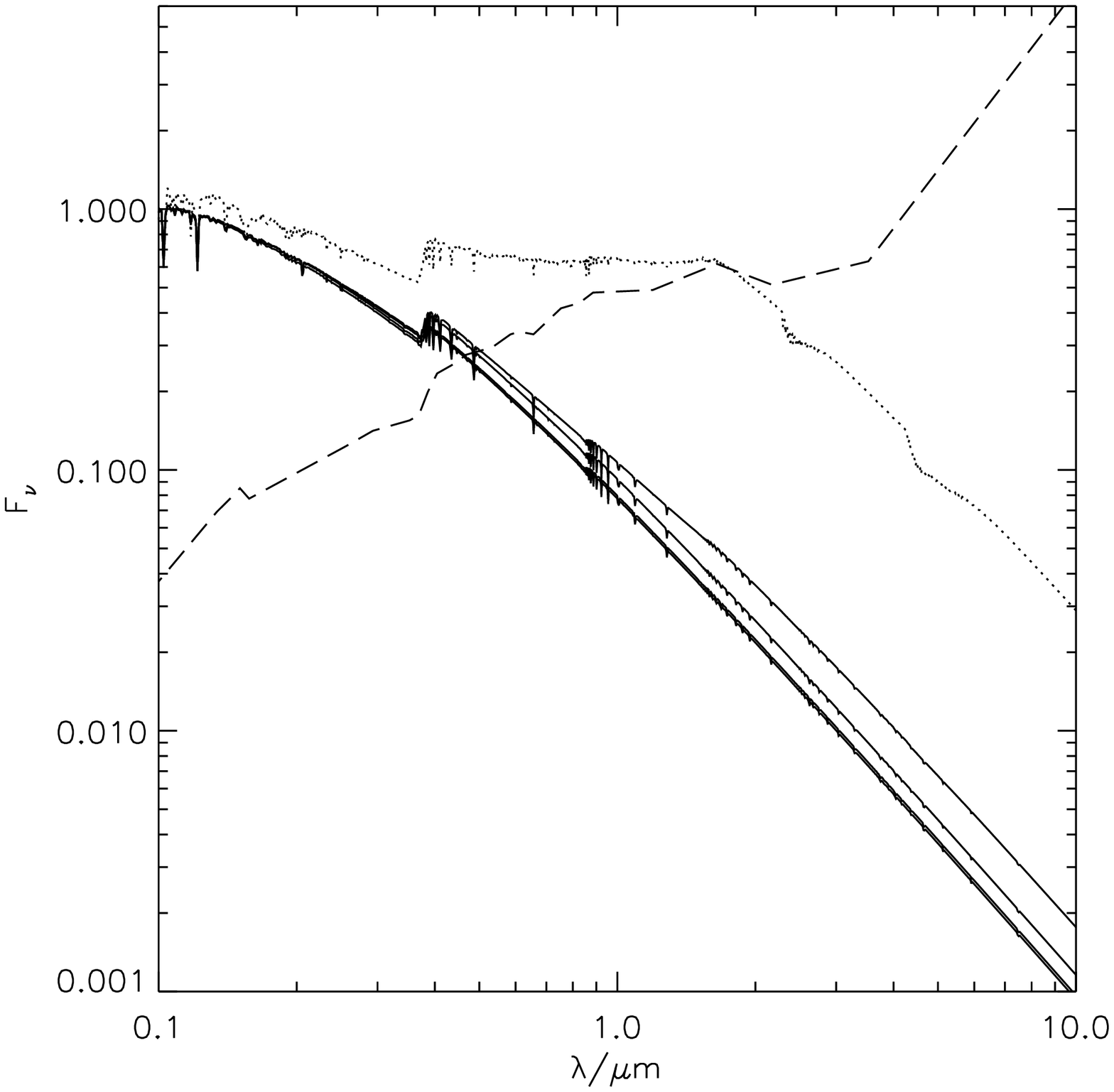,height=9cm,angle=0}}
{\bf Figure 1.} Spectral energy distribution for a starburst with constant star
formation rate, age of 10 million years, metallicity $Z=0.01 Z_{\odot}$ and a
Salpeter ($x=-1.35$) IMF with four different lower mass cutoffs: 0.1, 2.0, 5.0
and 10.0 M$_{\odot}$ (solid lines), from top to bottom.  Starbursts with no
low mass stars ($< 5$ M$_{\odot}$) are about a factor of 2 fainter than
starbursts with low mass stars ($> 0.1$ M$_{\odot}$), at 3 $\mu$m. Note that
for cutoff masses larger than 5 M$_{\odot}$, the spectral energy distribution
changes very little, making our predictions almost independent of the exact
value of the cutoff mass. The models have been normalised at 1000 \AA. Also
shown is a solar metallicity starburst from \scite{SKCS97} (dashed line). We
finally plot an idealized starburst model with constant star formation rate,
Salpeter IMF, age of 10 million years, $Z=0.2 Z_{\odot}$ and no dust (dotted
line). The UV SED of this starburst can be a rough approximation of the UV SED
of quasars.
\end{figure}

We have computed the SED of PPGs using the extensive set of synthetic stellar
population models developed in \scite{JPMH98,J+99}. In particular, PPGs are
modelled using a constant star formation rate and $Z=0.01 Z_{\odot}$
metallicity. As noted above, we expect (and therefore assume) that no
photometrically significant quantity of dust has already been formed in
PPGs. We also assume that PPGs are young objects (age less or equal than 100
million years). Furthermore, and in order to test our primordial IMF
hypothesis -- see \S 4, we have adopted a Salpeter IMF ($x=-1.35$) with four
different low mass cutoff values: 0.1, 2, 5 and 10 M$_{\odot}$ (i.e. the IMF
does not contain any stars with masses below the cut-off values).

The effect of the different low mass cutoff values on the SED of a PPG is
shown in Figure~1, where the flux density F$_{\nu}$ is plotted in arbitrary
units for a 10 million year old PPG with $Z=0.01 Z_{\odot}$ and constant star
formation (solid lines).  Different solid lines corresponds to the different
IMF cutoffs, 0.1, 2, 5 and 10 M$_{\odot}$, from top to bottom (the 5 and 10
M$_{\odot}$ are almost overlapped). As expected, the lack of low mass stars
translates into a deficit of flux at IR wavelengths. Furthermore, for a low
mass cutoff larger than 5 M$_{\odot}$ the SEDs do not differ much since stars
with masses larger than 5 M$_{\odot}$ have similar spectral energy
distributions in the IR. The difference in relative flux density over the wide
wavelength range illustrated in Figure~1 provides the means to test the IMF in
PPGs -- see \S 4.  The SED of an observed nearby starburst, from
\scite{SKCS97}, is also plotted in Figure~1 (dashed line), together with a 10
million year model starburst, with continuous star formation, metallicity
$Z=0.2 Z_{\odot}$, and no dust extinction or emission (dotted line).  The
difference in the spectral slope between PPGs and nearby starbursts is
striking, and it is still significant between PPGs and the idealized starburst
model with no dust. Although the idealized starburst model with no dust is
unlikely to describe any galaxy observed nearby, it can be used as an
approximate model for the UV SED of quasars, due to its rather flat SED in the
far ultra violet wavelengths.

We have computed AB magnitudes ($m_{\rm AB}=-2.5{\rm log}(F_{\nu})-48.59$,
where $F_{\nu}$ is expressed in erg s$^{-1}$ Hz$^{-1}$) for 3 different
wavelengths: 1.2, 3.6 and 8 $\mu$m. Figure~2 shows the colour--colour
trajectory for PPGs with a Salpeter IMF with a low mass cutoff of 0.1
M$_{\odot}$ (thin solid line), and 5 M$_{\odot}$ (thick solid line),
metallicity $Z=0.01 Z_{\odot}$, and an age of 10 million years.  The dashed
line is the evolution of the nearby starburst from Figure~1, and the dotted
line the evolution of the idealized starburst model with no dust and
$Z=0.2Z_{\odot}$, also from Figure~1.  All models have been simply
K--corrected and the numbers that label the trajectories indicate the
corresponding redshift. The figure shows that PPGs with redshift range
$5<z<10$ have colors in the range of values
$-3.0\le(1.2\mu$m$-8\mu$m$)_{AB}\le-2.0$ and
$-0.9\le(3.6\mu$m$-8\mu$m$)_{AB}\le-1.5$, that are inside the dashed area. The
idealized starburst model with no dust and sub--solar metallicity enters
marginally the dashed area, but only for low redshifts ($z<0.5$). This
idealized model is unlikely to describe nearby galaxies, while it can be a
rough description of the UV SED of quasars, in which case it could be
concluded that PPGs should have very different colors than quasars at redshift
$z>0.5$ (at least about 0.5~mag away in the color--color plot of Figure~2).
Nearby galaxies are either star forming galaxies with significant amount of
dust (irregular, spiral or starburst galaxies), or older stellar systems with
little gas or dust (elliptical galaxies). In both cases nearby galaxies are
much redder than PPGs. In Figure~2, the dashed dotted lines show the
color--color redshift evolution of two typical nearby galaxies (a spiral and
an elliptical galaxies, from \scite{SKCS97}). These galaxies, even if nearby,
are always at least 2~mag redder than PPGs in the $(1.2\mu$m$-8\mu$m$)_{AB}$
color.

The fact that PPGs are the bluest objects, in this IR color--color plot, is an
important result, since a photometric search for high redshift galaxies would,
in principle, be biased towards selecting the solar metallicity starbursts,
which are the reddest galaxies, as already proposed by \scite{SE99}. Although
it cannot be excluded that galaxies of metallicity close to the solar value
might exist at $z=10$, and that PPGs are rare (since they are young by
definition), Figure~2 shows that primordial star formation at high redshift
should be searched for in very blue objects. The shaded area in Figure~2 marks
the expected location of PPGs.  Even if young stellar populations of
intermediate metallicity ($0.01Z_{\odot}\le Z\le 0.1Z_{\odot}$) and low
redshift ($z<1$) have colours inside the shaded area of Fig.~2, it is likely
that real star forming galaxies of that metallicity and redshift are
significantly reddened by dust (emission and absorption). As an example, for a
$z=1$ galaxy with $Z=0.1Z_{\odot}$ and dust the 8 $\mu$m corresponds to 4
$\mu$m, where thermal emission from dust can be appreciable (see Figure~1). If
galaxies with no dust and intermediate metallicity ($0.01Z_{\odot}\le Z\le
0.1Z_{\odot}$) and redshift ($1<z<3$) exist, they are likely to contain no
massive stars, since star formation must be finished (no gas left), and must
also have much redder colours than PPGs.

\begin{figure}
\centerline{ \psfig{figure=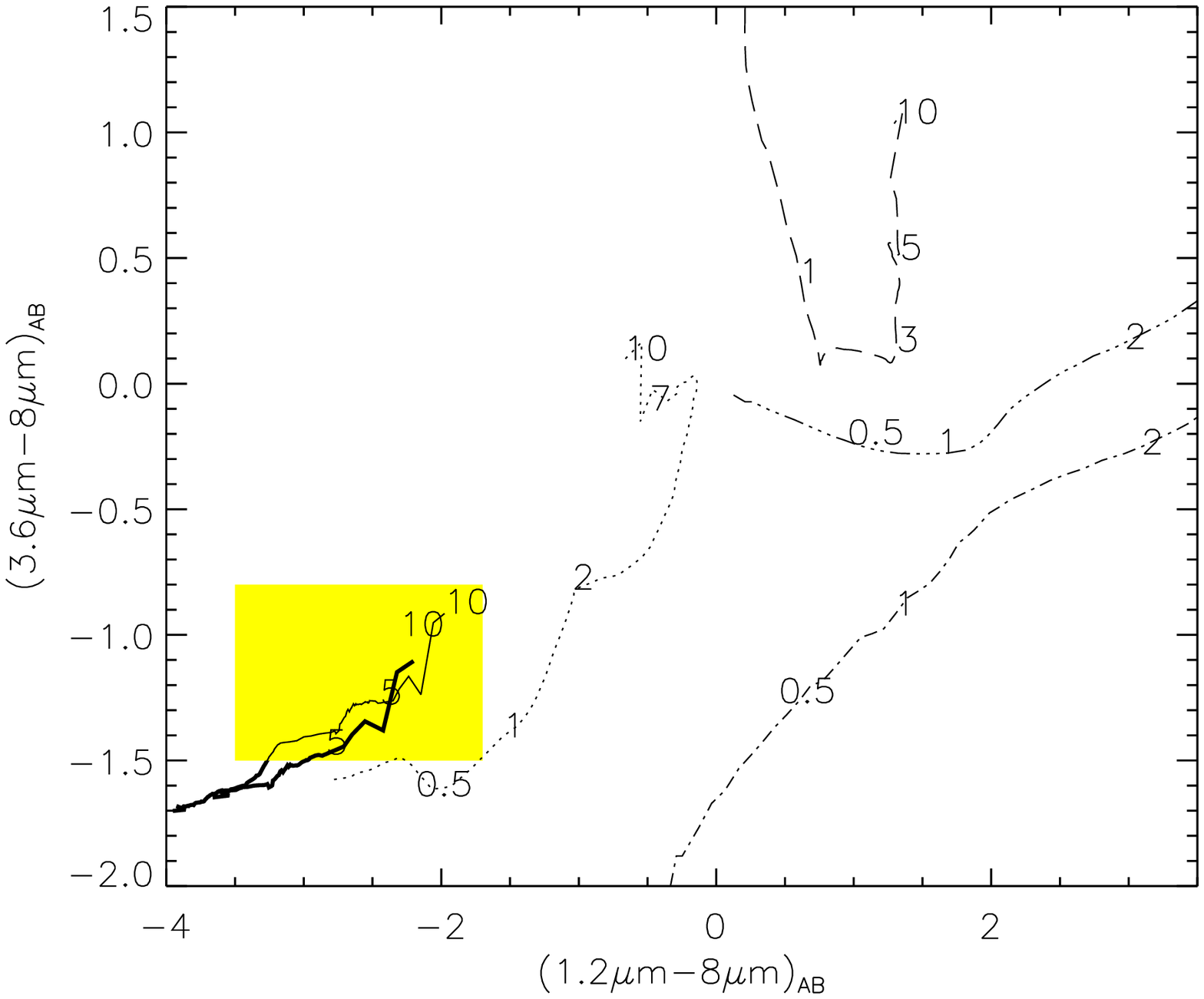,height=7cm,angle=0}} 
{\bf Figure 2.} Colour--colour (in the AB system) redshift evolution of: the 
observed
starburst of solar metallicity from Figure~1 (dashed line); the idealized
starburst model of 1/5 solar metallicity and no dust, also from Figure~1
(dotted line); two PPGs models (from Figure~1), with constant star formation
rate, age of 0.01~Gyr, $Z=0.01 Z_{\odot}$, and Salpeter IMF with
0.1~M$_{\odot}$ cutoff (thin solid line) and 5~M$_{\odot}$ cutoff (thick solid
line); two typical nearby galaxies (a spiral and an elliptical), from
\scite{SKCS97}. The hatched area shows where PPGs are expected to be.
\end{figure}

\section{The IR Flux Density of PPGs}

In the previous section it has been shown that PPGs can be photometrically
selected as the bluest galaxies in the Universe.  The question that needs to
be answered now is: Can PPGs be detected at all with future telescopes such as
the SIRTF and the NGST? PPGs could in fact be very faint because they could be
very small, or because their star formation rate (SFR) could be very low (see
\S 6).  To answer this question, we have computed the expected flux in nJy per
unit of SFR (in M$_{\odot}/$yr) and per unit of wavelength (in $\mu$m), as a
function of redshift, for a PPG with a Salpeter IMF and cutoff mass of 5
M$_{\odot}$, in an Einstein--deSitter Universe ($H_0=65$ km s$^-1$ Mpc$^{-1}$)
and in an open Universe ($\Omega_0=0.2$). The plots are roughly independent of
the starburst age, for age larger than a few million years.  Figure~3 shows
that, although PPGs do not exhibit a negative K-correction as galaxies do in
sub-mm and mm bands, they suffer from very little cosmological dimming, even
in an open Universe. From Figure~3, it can be seen that PPGs with SFR of about
$100$ M$_{\odot}/$yr have a flux in the faintest band ($8\mu$m) of about 1 nJy
(although they would be much brighter at 1.2 $\mu$m -- about 1 $\mu$Jy). The
IRAC camera at SIRTF will be able to measure in 1 hour exposure a flux of
5$\times 100$~nJy at the 5$\sigma$ level, and will not be able to detect
PPGs. On the other hand, the NGST will be able to achieve 1 nJy within 50
hours of observation ({\it www.ngst.stsci.edu}). It does then seem plausible
that the NGST will be able to carry a deep enough survey to detect PPGs with a
SFR of at least $100$ M$_{\odot}/$yr.

\begin{figure}
\centerline{ \psfig{figure=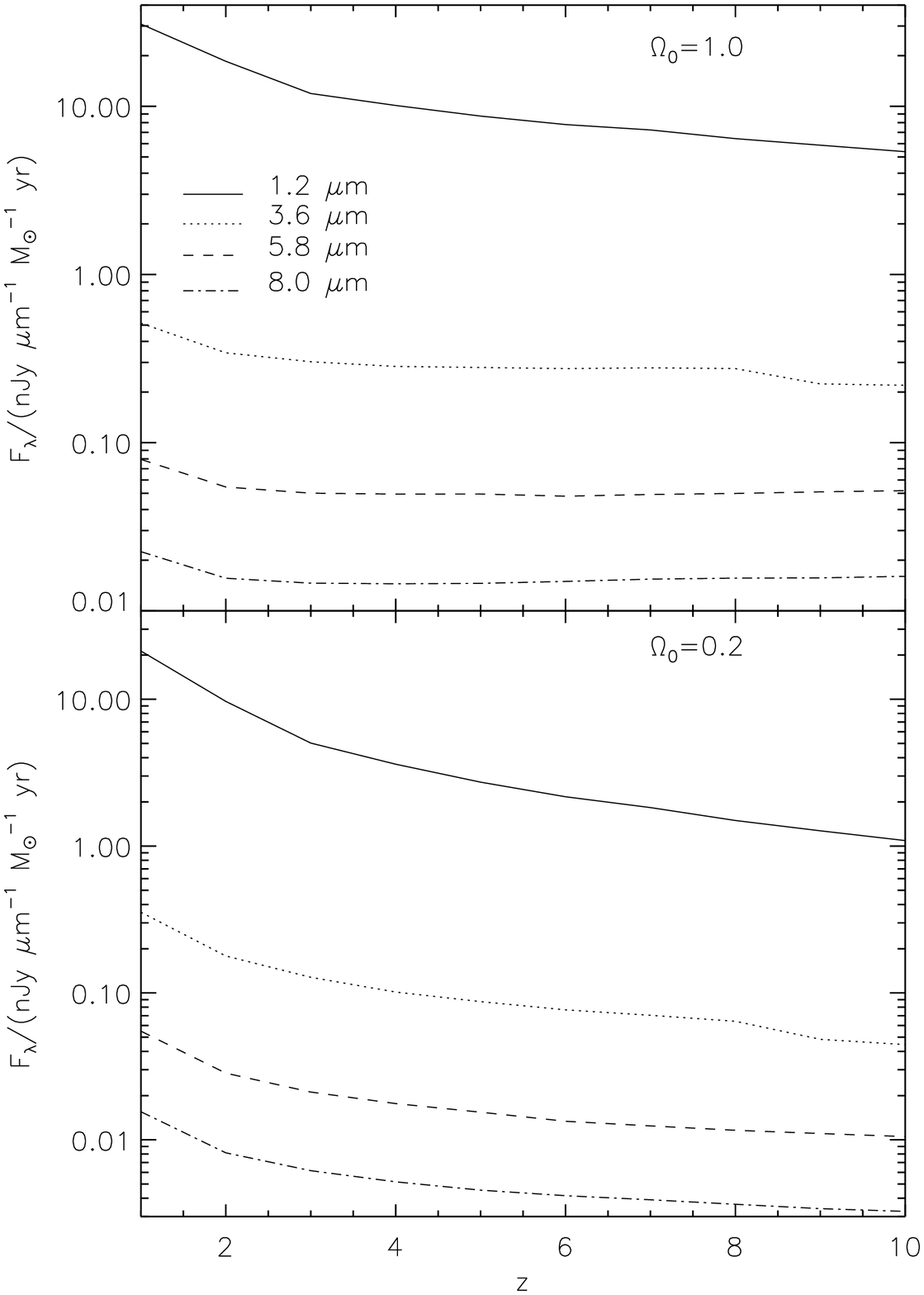,height=12cm,angle=0}}
{\bf Figure 3.} The expected IR flux density of PPGs, per unit of star 
formation
rate, as a function of redshift, for four wavelengths and 2 cosmologies (in
both cases $H_0=65$ km s$^-1$ Mpc$^{-1}$). Note that the cosmological dimming
is very small since PPGs have $F_{\nu} \propto \nu^{1.5}$ (see Figure~1). The
same model as in Figures~1 and 2 has been used, with a 5~M$_{\odot}$ cutoff
IMF. The flux density has been computed for an age of 0.01~Gyr, and increases
rather slowly afterwards.
\end{figure}

\section{The Primordial Stellar IMF}

Observational and theoretical arguments suggest that stars forming from gas of
zero metallicity could have an initial mass function (IMF) shifted towards
much larger masses than stars formed later on from chemically enriched gas.
The observational arguments are extensively discussed in a recent paper by
\scite{L98}, and we refer the reader to that work. The main theoretical reason
in favor of a `massive' zero metallicity IMF is the relatively high
temperature ($T>100$~K) of gas of primordial composition, that is cooled at
the lowest temperatures mainly by H$_2$ molecules
(\pcite{PSS83,MS86,SK87,KSFR90,KS92,AN96}).

One of the most exciting aspects of the photometric discovery of PPGs would be
the possibility of investigating the nature of their stellar IMF, once their
redshifts will be available. The knowledge of the zero metallicity IMF is
fundamental for modeling the early chemical evolution of galaxies, and for
estimating the Population III ionizing radiation field in the Universe
(\pcite{HL97,F98}). Because there is no definitive theory for the origin of
the stellar IMF, and due to the difficulty of modeling or probing directly the
thermo--dynamic state of protogalactic gas of primordial composition, the
issue of the zero metallicity IMF is still open.

The stellar IMFs estimated in different clusters, associations, and for the
solar neighborhood are in general consistent with a power law with Salpeter
exponent, $x=1.35$, for stellar masses larger than about $1$~M$_{\odot}$
(\pcite{E98}), although exceptions exist (\pcite{S98}). The power law shape of
the stellar IMF and the similarity of the IMFs emerging from different star
formation sites suggest that the stellar IMF is the result of a scale free
dynamical process, not very sensitive to local properties of the ISM such as
chemical compositions, temperature, density, etc. However, the power law shape
of the IMF, and so the self similarity of the dynamical process at its origin,
must be broken at some small scale, because only a finite fraction of the dark
matter in stellar clusters or in the galactic halo and disk systems can be
made of brown dwarfs (\pcite{GBF97,RG97,K97,GFB98,R+99}). It is likely that
while the local properties of the ISM do not interfere significantly with the
self similar dynamics that originate the stellar IMF, they do play a role in
setting the particular value of the mass scale where the self similarity is
broken.  According to this point of view, one expects the stellar IMF to have
always more or less the same power law shape, down to a cutoff mass whose
value depends on local properties of the ISM, and up to the largest stellar
mass, whose value is limited either by the total mass of the star formation
site, or by some physical process that prevents the formation of
super--massive stars.

The lower mass cutoff of the stellar IMF has been predicted in models of i)
opacity limited gravitational fragmentation
(\pcite{S177,S277,S377,YS85,YS86}); ii) protostellar winds that would stop the
mass accretion onto the protostar (\pcite{AF96}); iii) fractal mass
distribution with fragmentation down to one Jeans' mass (\pcite{L92}).  If
gravitational sub--fragmentation during collapse is not very efficient (see
\scite{B93}), the value of the Jeans' mass determines the lower mass cutoff of
the IMF. In \scite{PNJ97}, numerical simulations of super--sonic and
super--Alfv\'{e}nic (\pcite{PN99}) magneto--hydrodynamic turbulence are used
to compute the probability density function of the gas density, which is used
to predict the distribution of the Jeans' mass in turbulent gas, under the
reasonable assumption of uniform kinetic temperature.  The Jeans' mass
distribution computed in \scite{PNJ97} has an exponential cutoff below a
certain mass value, that is found to be:
\begin{equation}
M_{\rm min}= 0.2M_{\odot}\left(\frac{n}{10^3cm^{-3}}\right)^{-1/2}
\left(\frac{T}{10 K}\right)^{2}\left(\frac{\sigma_{v}}{2.5 km/s}\right)^{-1}
\label{1}
\end{equation}
where $T$ is the gas temperature, $n$ the gas density, and $\sigma_{v}$ the
gas velocity dispersion. Using the ISM scaling laws, according to which
$n^{1/2}\sigma_{v}\approx const$, one obtains:
\begin{equation}
M_{\rm min}\approx 0.1 M_{\odot}\left(\frac{T}{10K}\right)^2
\label{2}
\end{equation}
that is a few times smaller that the average Jeans mass (the
Jeans mass corresponding to the average gas density), and therefore an
important correction to more simple models of gravitational fragmentation,
that do not take into account the effect of super--sonic turbulence on the gas
density distribution.

If the ISM has primordial chemical composition, and the main coolant is
molecular hydrogen, a temperature below $100$~K is hardly reached, and the
stellar IMF might have a lower mass cutoff of about $10$~M$_{\odot}$. Similar
lower mass cutoffs are obtained in the models by \scite{S377} and
\scite{YS86}, who estimated typical stellar masses, based on molecular
hydrogen cooling, of approximately $20$ and $10$~M$_{\odot}$ respectively.
More recent numerical simulations of the collapse and cooling of cosmological
density fluctuations of large amplitude (the first objects to collapse in the
Universe), yield even larger values of the Jeans' mass, of the order of
$100$~M$_{\odot}$ (Bromm, Coppi, \& Larson 1999; Abel, Bryan, \& Norman 1998).

The discovery of PPGs could shed new light on the problem of the primordial
IMF. The redshift evolution of the two colors $(1.2\mu$m$-8\mu$m$)_{AB}$ and
$(3.6\mu$m$-8\mu$m$)_{AB}$, computed with the PPG model discussed in this
work, is plotted in Figure~4.  The solid line is the case of a PPG with a
Salpeter IMF with $5$~M$_{\odot}$ cutoff, and the dashed line the same PPG
model, but with a $0.1$~M$_{\odot}$ cutoff. Once a PPG candidate is selected
with the IR broad band photometry as a very blue object (colors inside the
shaded area in Figure~2), and its redshift is estimated with a Lyman drop
method, or with an H$\alpha$ search, the IR colours provide a tool to
discriminate between a standard IMF, and an IMF deprived of low mass stars.
The lower panel of Figure~4 shows that a PPG with a 'massive' IMF can be about
0.5~mag bluer in $(1.2\mu$m$-8\mu$m$)_{AB}$ than a PPG with a standard IMF.

\begin{figure}
\centerline{ \psfig{figure=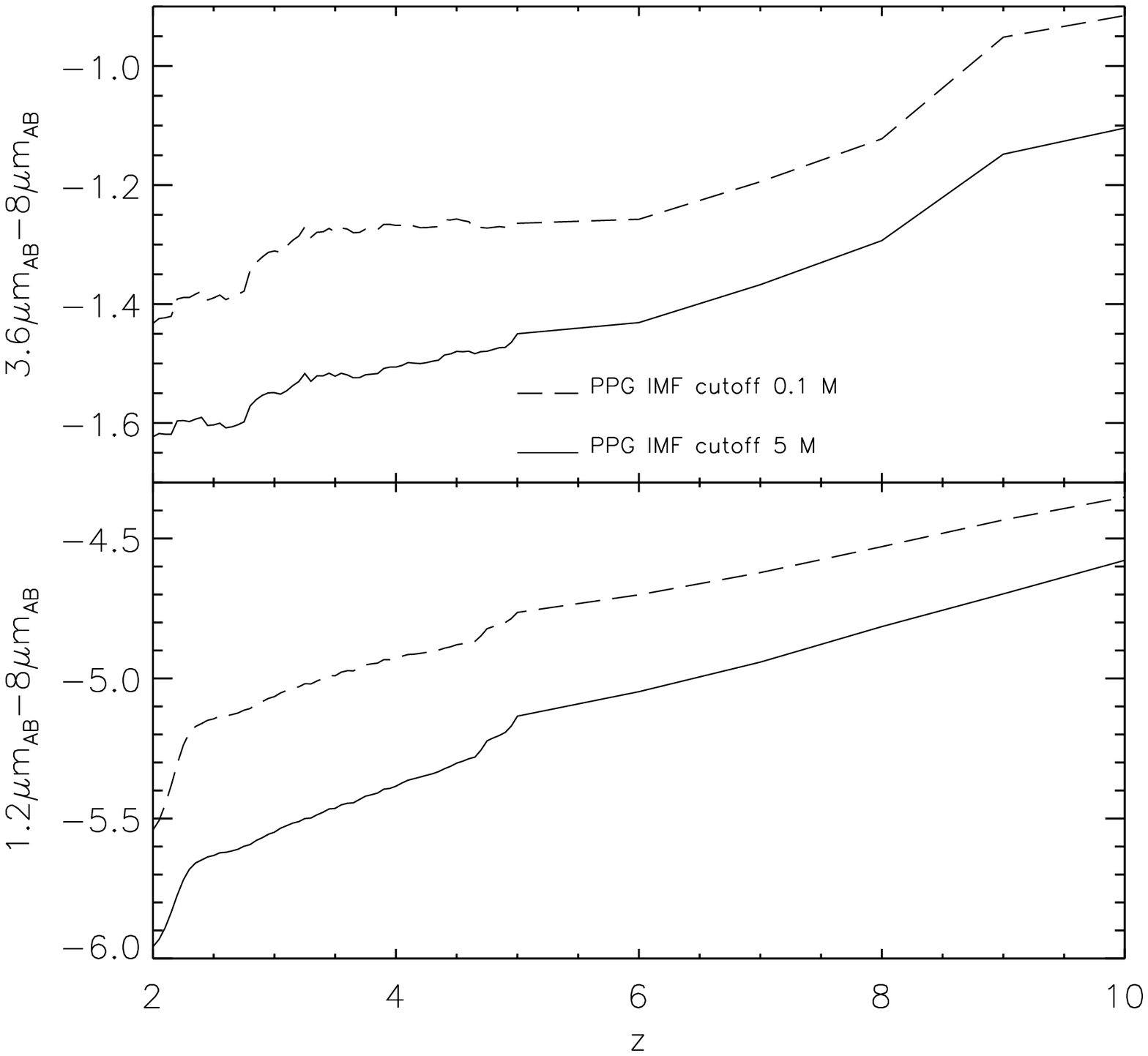,height=8cm,angle=0}}
{\bf Figure 4.} Colour evolution (AB system) with redshift for PPGs with a 
Salpeter
IMF with $0.1$~M$_{\odot}$ cutoff (dashed line), and $5$~M$_{\odot}$ cutoff
(solid line). As expected from Figure~1, the biggest difference occurs for
colours that sample both the near-IR (rest--frame UV) and the far-IR. The
difference between the Salpeter and the 'massive' IMF is about 0.5 mag in
$(1.2\mu$m$-8\mu$m$)_{AB}$. Once PPGs have been found at high $z$ using the
broad band color--color selection from Figure~2, and their redshift has been
determined with narrow band photometry, the $(1.2\mu$m$-8\mu$m$)_{AB}$ color
can be used to discriminate among objects with a standard Salpeter IMF or a
$5$~M$_{\odot}$ cutoff IMF.
\end{figure}

\section{Detectability via broad-band infrared imaging versus
emission-line searches}

In this work we propose to select PPGs as the bluest objects in deep IR
surveys, on the basis of the color--color plot shown in Figure~2. We now
address the question of how a broad band photometric selection of PPGs
performs, compared with searches of emission lines, such as Lyman-$\alpha$ and
H-$\alpha$.  The rest frame equivalent widths of Lyman-$\alpha$ and H-$\alpha$
can be very roughly estimated by assuming that each photon below 1251 and 1025
\AA\, will originate a Lyman-$\alpha$ and H-$\alpha$ photon respectively.  The
equivalent widths estimated in this way are of course upper limit to the true
equivalent widths.  We find that the equivalent width of Lyman-$\alpha$ is 380
\AA\, while the equivalent width of H$-\alpha$ is 4400 \AA\ --the latter is so
high due to the fact that the continuum at 6563 \AA\ is rather faint in PPGs
(see Fig.~1).  Assuming that the lines have intrinsic widths at rest--frame
typical of the virial velocity of a galaxy (for example a line width of
300~km/s corresponds to 2 and 13 \AA\ respectively), one finds that they will
only be about 30 times brighter than the continuum at $z \sim 10$. Since one
would need to shift the narrow filter for about 500 steps, or more, to search
for all possible emitters in the redshift range $5\le z\le 10$, the advantage
of the lines being brighter than the continuum is offset by the number of
steps needed to find all PPGs between $z=5$ and 10. It seems therefore that IR
broad band photometry is an easier way to both detect and select PPG
candidates than the emission line technique, because only one deep exposure is
needed to find {\it all} PPGs in the redshift range $5\le z\le 10$. Note that
the equivalent width of Lyman-$\alpha$ and H-$\alpha$ has been over--estimated
here.  Moreover, the advantage of deep broad band photometry is that, together
with detecting and selecting PPGs, it provides at the same time important
information about their stellar populations.  However, it is important that
the emission line technique (or a Lyman break technique) is available aboard
the NGST, since photometric redshifts measured with narrow filters will
probably be the best (or the only) way to further constrain the redshift of
PPG broad band photometric candidates, which is necessary to extract
information about their stellar population from the broad band colors
(Figure~4).

\section{Discussion and Conclusions}

We have studied the photometric properties of very young proto--galaxies with
primordial or very low ($Z=0.01 Z_{\odot}$) metallicity and no significant
effect of dust in their SED. We have named these galaxies ``primordial
protogalaxies'', or PPGs. Using the methods of synthetic stellar populations,
we have shown that PPGs are the bluest stellar systems in the Universe. They
can therefore be selected in color--color diagram obtained with deep broad
band IR surveys, and can be detected with the NGST, if they have a SFR of at
least $100$~M$_{\odot}/$yr, over a few million years.  We have discussed the
possibility of using the IR colours of PPGs to constrain their stellar IMF,
and investigate the possibility that the stellar IMF arising from gas of
primordial chemical composition is more ``massive'' than the standard Salpeter
IMF. Finally we have argued that broad band photometry can be more convenient
than emission line searches, to detect and select PPGs.

It is possible that PPGs are rare because the chemical self--enrichment of a
proto--galaxy could be very fast and efficient (\pcite{PJJ97}), or that they
are difficult to detect, because population III stars could be formed at very
large redshift ($z>10$), in object of very low mass (e.g. \scite{HTL96}).  In
order to enrich gas of primordial composition to a metallicity $Z=0.01
Z_{\odot}$, with a standard Salpeter IMF, it is necessary to convert into
stars about 0.2\% of the gas mass.  A star formation rate of
$100$~M$_{\odot}/$yr over a few million years is necessary for detecting a PPG
with the NGST. With such SFR, after $10^7$ years $10^9$~M$_{\odot}$ of gas is
turned into stars. In order to still have a metallicity of $Z\le 0.01
Z_{\odot}$, these stars must be formed in a system with baryonic mass of at
least $1\times 10^{11}$~M$_{\odot}$, that is a very large galaxy, or a small
group of galaxies. Such massive systems are inside large dark matter halos
that are not collapsing yet at redshift $5\le z\le 10$. It is possible that
PPGs that can be detected with the NGST are the progenitors of very large
galaxies, in a phase when their dark matter halo has not turned around yet.
If PPGs are discovered, their spectro--photometric properties could give very
important clues for the problem of star formation in galaxies (such as the
origin of the stellar IMF) and their luminosity, abundance, and redshift
distribution would trace the complete history of the very first star formation
sites in the Universe.


\end{document}